  \documentclass[structabstract]{aa}  
\usepackage{natbib}
\bibpunct{(}{)}{;}{a}{}{,}

\usepackage{graphicx}
\usepackage{txfonts}
\begin{document}
   \title{Center-to-limb variation of the area covered
        by magnetic bright points in the quiet Sun}
   \author{J.~A. Bonet\inst{1,2} \and 
        I.~Cabello\inst{3} \and
        J.~S\'anchez Almeida\inst{1,2} 
          }

   \institute{Instituto de Astrof\'\i sica de Canarias, E-38205 La Laguna,
    Tenerife,Spain\\
     \email{jab@iac.es}
      \and 
     Departamento de Astrof\'\i sica, Universidad de La Laguna,
     Tenerife, Spain 
     \and
                Image Processing Laboratory, Universidad
                de Valencia, E-46980 Paterna, Valencia, Spain\\
             }

   \date{Received September 15, 1996; accepted March 16, 1997}

 
\abstract
   {The quiet Sun magnetic fields produce ubiquitous bright points (BPs)
        that cover a significant fraction of the solar surface.
        Their contribution to the total solar irradiance (TSI)
        is so-far unknown.}
   {To measure the center-to-limb variation (CLV) of the fraction
        of solar surface covered by quiet Sun magnetic bright 
        points.
The fraction is referred to as {\em fraction of covered surface}, or FCS.
        }
   {Counting of the area covered by BPs in $G$-band images 
        obtained at various heliocentric angles with
        the 1-m Swedish Solar Telescope on La Palma. 
        Through restoration, the images are close to the 
        diffraction limit of the instrument 
        ($\sim 0\farcs 1$).
        }
   {
The FCS  is largest at disk center ($\simeq$ 1\,\%),
and then drops down to become $\simeq$ 0.2\,\% at 
$\mu\simeq 0.3$ 
(with $\mu$ the cosine of the heliocentric angle).
The relationship has large scatter, which we 
evaluate comparing different subfields within our FOVs.
We work out a toy-model to describe the 
observed CLV, which considers the BPs to be 
depressions in the mean solar photosphere
characterized by a depth, a width, and a spread of 
inclinations. Although the model is poorly constrained by observations, 
it shows the BPs to be shallow structures
(depth$\,<\,$width) with a large range of inclinations.
We also estimate how different parts of the solar disk
may contribute to TSI variations, finding that 
90\,\% is contributed by BPs having $\mu > 0.5$,  
and half of it is due to BPs with $\mu > 0.8$. 
   }
   {}


\maketitle

\section{Introduction}\label{intro}

Our understanding of the quiet Sun magnetic fields
has drastically improved during the last decade
\citep[for recent reviews, see,
e.g.,][]{2009SSRv..144..275D,2011ASPC..437..451S}.
We have gone from 
magnetic signals present only at the network boundaries 
\citep[e.g.,][]{bec77b,sol93}, to ubiquitous 
polarization signals created through Hanle effect
\citep[e.g.,][]{fau93,tru04} and Zeeman effect   
\citep[e.g.,][]{lin99,2000ApJ...532.1215S,2003ApJ...582L..55D,har07,lit08}.
The wealth of quiet Sun magnetic structures 
makes them  potentially important to 
understand the global magnetic properties of the Sun 
\citep[][]{2003ApJ...585..536S,tru04}, 
and also makes it unlikely that the quiet Sun magnetism 
results from the decay of active regions 
\citep[e.g.,][]{san03b}.
Theoretical arguments, corroborated by numerical experiments, 
favor a different production mechanism 
\citep{pet93,cat99b,vog07,pie10,2011ApJ...736...36M}. 
An efficient turbulent dynamo transforms into magnetic fields
part of the kinetic energy of the granular convection. 
It generates a complex magnetic field which evolves in 
short time scales (a few min) and 
has small characteristic length-scales ($<\,1\,$Mm).

In agreement with the turbulent dynamo scenario,
quiet Sun magnetic fields come with strengths 
in the full range covering from almost zero to 2\,kG 
\citep[][]{2000ApJ...532.1215S,dom06,2008A&A...477..953M,
2009A&A...506.1415B,2011A&A...526A..60V}. Even if they 
only fill a small fraction of the quiet photosphere,
the part having strong kG fields may be particularly 
important for a number of reasons. 
Firstly, the magnetic flux and 
energy increase with field strength, therefore, 
the energy and flux provided by kGs may surpass 
the contribution of the more  common but weaker fields 
\citep[][]{san04}. The need to consider kGs
is illustrated by the numerical experiments
set up by \citet{2011arXiv1108.1155C}, where realistic
granular convection redistribute initial hG fields 
so that daG, hG and kG field strengths have the same 
energy despite their very different area covering.
Magnetic concentrations with kG fields may 
also be  important because  buoyancy makes them 
vertical \citep[e.g.,][]{sch86} and so, they naturally
provide  a mechanical connection between the photosphere and 
the upper atmosphere \citep[e.g.,][]{bal98,sch03b,goo04,jen06,san08}.
They can function as guides that sustain magneto-acustic wave
propagation, or be physical channels connecting plasmas 
of different atmospheric layers.
Finally, kG magnetic concentrations are expected to be 
particularly bright due to the 
so-called hot-wall effect\footnote{The 
magnetic pressure suffices to maintain 
kG structures
in mechanical balance within the photosphere,
therefore, they are evacuated and transparent, 
allowing us to look through into 
the sub-photosphere, which is generally hotter and so 
brighter.}\citep[]{spr77,car04,kel04}.
They produce bright points (BPs) which, depending
on the variation across the solar disk, 
and during the solar cycle, 
may even contribute to the Total Solar Irradiance (TSI) 
variations as network and plage magnetic fields do 
\citep[e.g.,][]{1997ARA&A..35...33L,
        2004A&ARv..12..273F,    
        2011SSRv..tmp..133F}.
        
The finding of BPs in the quiet Sun was immediately 
identified as the kG magnetic concentrations inferred 
from polarization measurements \citep{2004ApJ...609L..91S}. 
Their basic properties and their ubiquitous presence 
have been confirmed by a number of researchers 
\citep[][]{dew05,dew08,bov08,san08,
        2009ApJ...700L.145V,2010ApJ...723..787V,
        2010ApJ...714L..31G}.
They vary on time scales of minutes similar to that of 
granulation, and many BPs are at the resolution limit of the 
current instrumentation (some 0.1~Mm). They are extremely
common, filling at least 1\% of the solar surface
and outnumbering granules
\citep{2010ApJ...715L..26S}.    

\citet{2011A&A...532A.136S} have recently estimated
the excess of brightness produced by the quiet 
magnetic fields at the  disk center, turning out to be of the order 
of 0.15\,\%.  Their contribution is larger than the 
0.08\,\%~TSI variations associated with cycle, however, 
determining the impact of these kG fields on TSI demands 
knowing their center-to-limb variation (CLV), as well as 
the variation of the quiet Sun magnetic fields with the solar 
cycle. These two properties, central to assess the role of quiet Sun 
fields on TSI, are  poorly known. As far as the variation along 
the cycle is concerned, the claims in the literature are for 
little variation  if any \citep{san03d,shc03,har10}. 
Unfortunately, the  uncertainty
of such claims can be as large as a factor two 
\citep{fau01}. 
There are several works on the 
CLV 
signals associated with the quiet Sun magnetic fields 
\citep[e.g.,][]{2008A&A...477..953M,lit08,
 2011ApJ...737...52L} however, to the best of our knowledge, 
nothing is known on the 
CLV 
of the quiet Sun magnetic 
BPs. This is precisely the subject of our work, i.e., 
providing a first observational description 
of the 
CLV 
of the quiet Sun BPs. To be 
more exact, we evaluate how the area covered by quiet Sun 
BPs varies with the position on the disk. 

Two main difficulties hinder the analysis. 
First, one has to use images with enough spatial 
resolution and of uniform quality, since the number of 
BPs depends critically  on the spatial resolution of 
the observations \citep{1996ApJ...463..797T,2010ApJ...715L..26S}.
This issue is sorted out using data from the 1-m Swedish Solar 
Telescope 
\citep[SST,][]{2003SPIE.4853..341S} obtained in a single day 
during moments 
of excellent seeing, and then restored to get images 
with uniform resolution close to the diffraction
limit of the instrument \citep{2005SoPh..228..191V}.
Second, the BPs can be misidentified with granule 
borders and other structures 
\citep[e.g.,][]{2007SoPh..243..121B}, a problem
particularly severe in near-limb images.
This second problem is addressed resorting to  the 
cumbersome method of eye-ball 
identification which, however, has been proven to be 
reliable for our purpose 
\cite[][and references therein]{2010ApJ...715L..26S}.

The work is organized as follows; the observations are 
presented in \S~\ref{observation}. The actual CLV 
measurements are described in \S~\ref{sec_ctl}. In principle,
such measurements has to be interpreted using realistic
models of magneto-convection  
(such as those used for plages by
\citeauthor{car04}~\citeyear{car04}
or 
\citeauthor{kel04}~\citeyear{kel04}),
where the 3-dimensional geometry of the photosphere
is considered consistently. 
It would require repeating the analyses carried out on the 
observed images using synthetic images from the
simulations. Then the CLV coming from a comprehensive battery of 
numerical simulations should be compared with the 
observations. This detailed realistic approach clearly 
exceeds the scope of the work. However, we attempt a
toy-modeling which, despite its simplicity, considers the 
key ingredients that within the hot-wall paradigm 
determine the CLV  (\S~\ref{interpretation}). 
The model is compared with the observed CLV in \S~\ref{results},
which constraints some of the properties of the magnetic 
structures. 
The implications for TSI variations are analyzed in \S~\ref{tsi_var}.
These results are discussed and put into context
in  \S~\ref{conclusions}.

%
%

\section{Observations and data reduction}\label{observation}

\begin{figure}
   \includegraphics[scale=0.7]{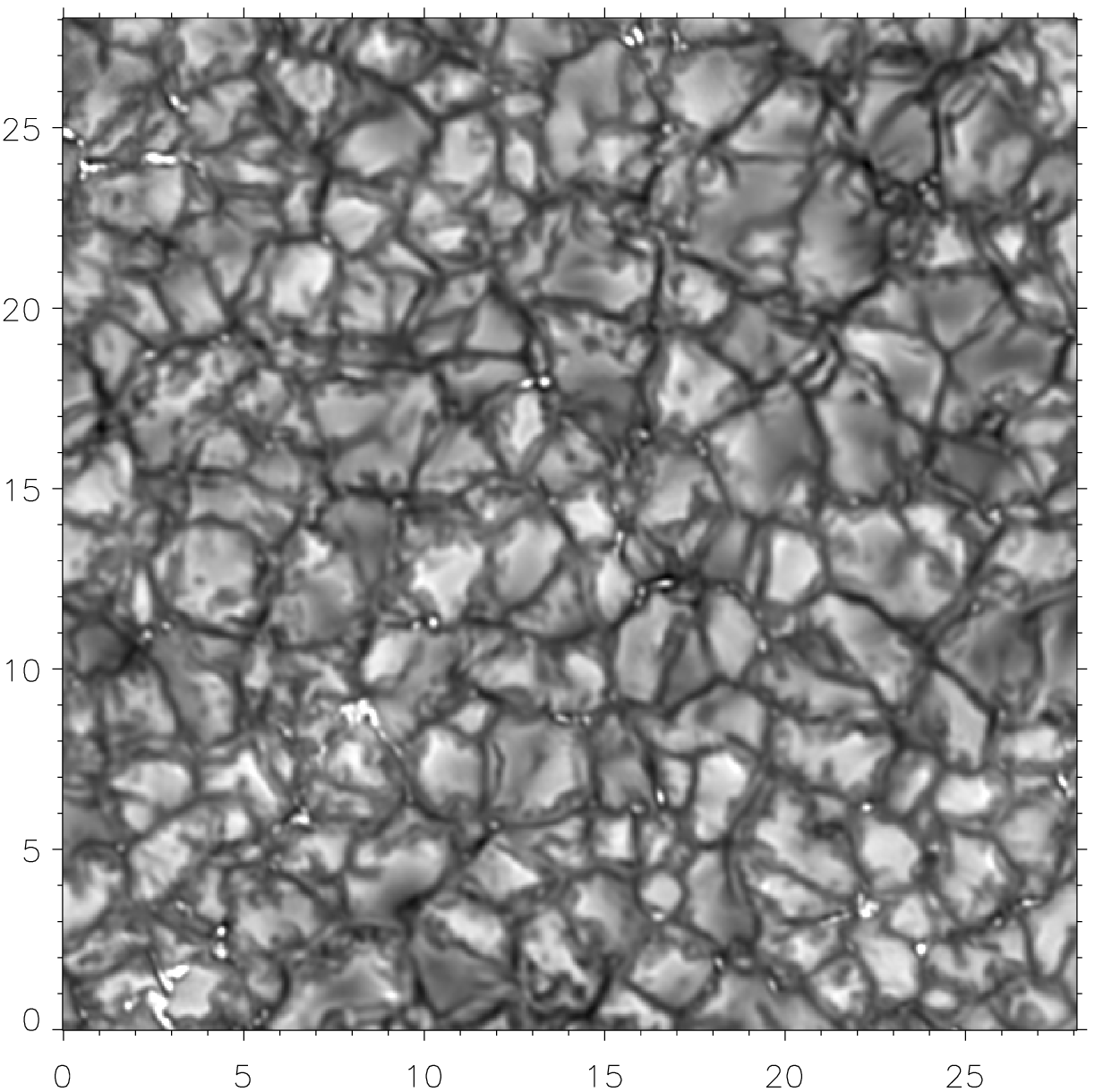}
   \includegraphics[scale=0.7]{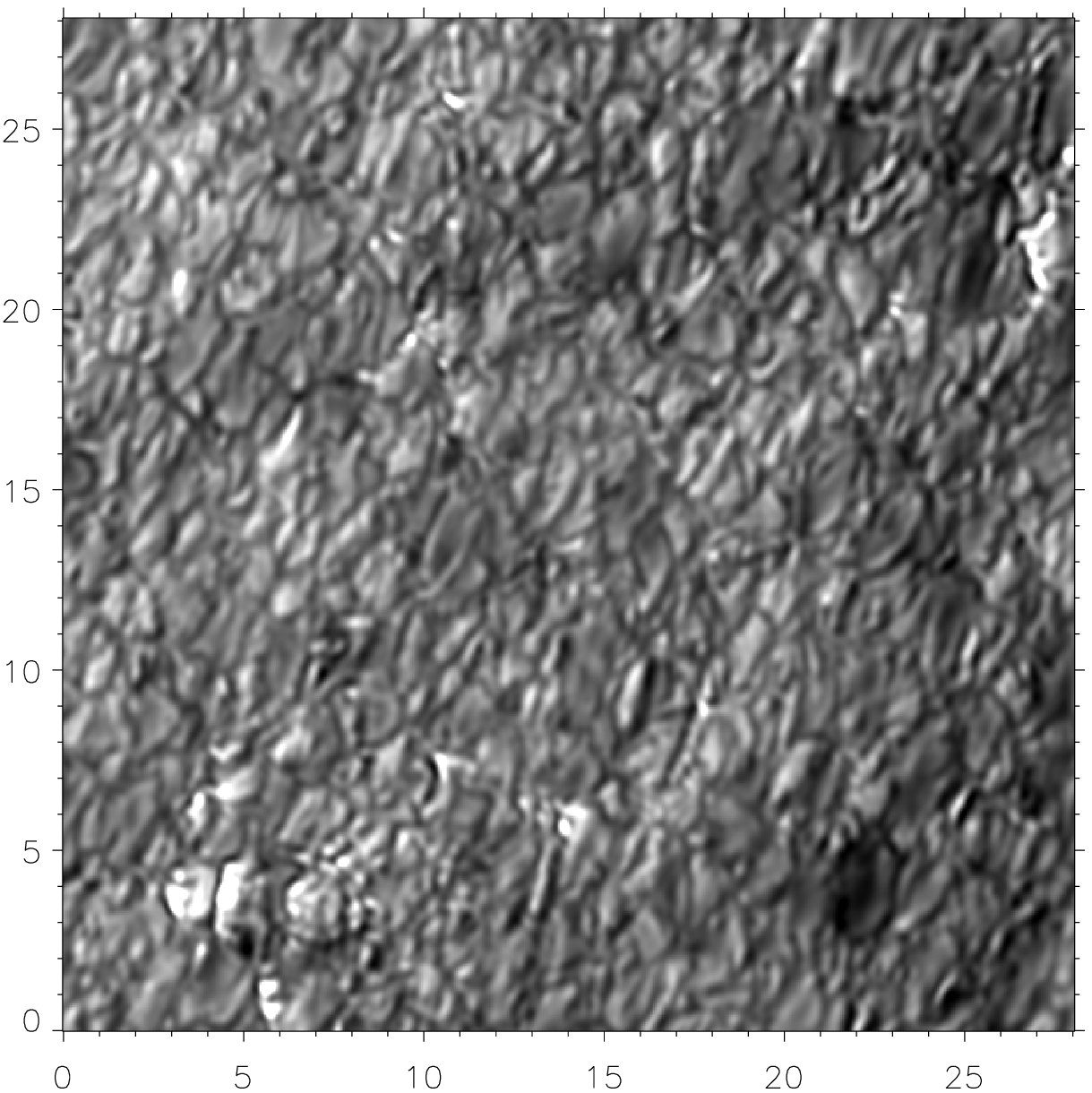}
\caption{Images to illustrate how the quiet Sun $G$-band
BPs look at different heliocentric angles -- 
top $\mu=0.99$, bottom  $\mu=0.53$. They correspond to the 
reference images in the series number 6 and 8, 
respectively (Table~\ref{ff_results}).
Axes represent arcsec from the lower left 
corner. 
        }
        \label{image}%
\end{figure}
Quiet Sun inter-network regions away from active areas
were observed on August 7, 2006,  
with the 1-m Swedish Solar Telescope 
\citep[SST, Roque de los Muchachos Observatory, La Palma;][]
{2003SPIE.4853..341S,2002Natur.420..151S}. 
The data consist of time series of images taken at eight 
different heliocentric angles, $\theta$, with $\mu(=\cos\theta)$
ranging from 0.34 to 1. The logbook,  
in Table~\ref{ff_results}, includes the heliocentric 
angle of each series as well as the UT of observation.
\begin{table*}                 
\caption{Description of the data sets including the
fraction of covered surface FCS.
}
\label{ff_results}
\centering
\begin{tabular}{c c c c c c c c}
\hline\hline
\noalign{ \smallskip}
Series & Initial time & Final time & Duration & \# of images & $\mu$ & FCS [$\%$] 
& $C^{\mathrm{a}}$ \\
\hline
\noalign{\smallskip}
1 & 8:19:00 & 8:35:45 & 0:16:45 & 100 & $0.999\pm0.002$ & $0.84\pm0.14$ &$1.04\pm0.13$ \\
2 & 8:47:16 & 8:54:36 & 0:07:20 & 42 & $0.610\pm0.028$ & $0.36\pm0.13$  &$1.16\pm0.18$ \\
3 & 8:56:41 & 9:12:58 & 0:16:17 & 98 & $0.582\pm0.030$ & $0.28\pm0.13$  &$1.21\pm0.15$ \\
4 & 9:16:12 & 9:29:02 & 0:12:50 & 78 & $0.341\pm0.065$ & $0.19\pm0.07$  &$1.15\pm0.13$ \\
5 & 9:32:44 & 9:48:54 & 0:16:10 & 96 & $0.802\pm0.027$ & $0.83\pm0.20$  &$1.10\pm0.15$ \\
6 & 9:49:53 & 10:06:25 & 0:16:32 & 98 & $0.995\pm0.004$ & $0.88\pm0.10$ &$1.05\pm0.15$ \\
7 & 10:09:02 & 10:23:13 & 0:14:11 & 86 & $0.926\pm0.013$ & $0.58\pm0.15$&$1.13\pm0.16$ \\
8 & 10:27:03 & 10:31:50 & 0:04:47 & 28 & $0.527\pm0.043$ & $0.31\pm0.10$&$1.22\pm0.15$\\
\noalign{\smallskip}
\hline
\end{tabular}
\begin{list}{}{}
\item[$^{\mathrm{a}}$] Mean intensity and rms variation 
        relative to mean photosphere
\end{list}
\end{table*}
Seeing was variable, with periods of excellence. As we explain below, 
we select for analysis only the best snapshots.

Images were  simultaneously recorded in three channels, two of them in 
$G$-continuum ($\lambda$~4363.9\,\AA; FWHM 11\,\AA) 
yielding pairs of simultaneous in-focus and out-of-focus images 
for post-facto application of phase diversity  
\citep{1992JOSAA...9.1072P}. The third channel, 
in the $G$-band ($ \lambda$~4305.6\,\AA; FWHM 10.8 \AA), is the one
used for our analysis. Each camera continuously gathered images at a 
rate of some 100 min$^{-1}$.
An additional fourth channel was devoted to CaII~H 
($\lambda$~3968.5\,\AA; FWHM~1.1\,\AA), which was
used during observation to avoid  active regions 
in the field of view (FOV). 
We employed two 10-bit Kodak Mega Plus 1536 $ \times $ 1024 cameras
for $G$-continuum, a 12-bit Redlake Mega Plus~II 1536 $ \times $ 1024 camera for
$G$-band, and a 10-bit Kodak Mega Plus 2024 $ \times $ 2048 camera for CaII H. 
The image scale was $0\farcs0405\,$pix$^{-1}$ in all cases.

After dark-current subtraction, flatfielding, and elimination of spurious pixels
and borders, the image restoration in the first three channels was performed 
simultaneously 
employing the Multi-Object Multi-Frame Blind Deconvolution method 
\citep[MOMFBD;][]{2005SoPh..228..191V}.
Sets of $\sim$30 images per channel (i.e., a total of
$\sim 3 \times 30$ images) were combined to produce a single pair of
simultaneous $G$-band and $G$-continuum restored snapshots. Thus, we achieved
two time series of restored images with a cadence of $\sim$10 s, an effective
FOV of 58\farcs6 $\times$ 38\farcs6, and an angular resolution close to the
diffraction limit of the SST at the working wavelengths ($\sim$0\farcs1). The
images of these series were de-rotated to compensate diurnal field rotation,
rigid-aligned, destretched to remove image distortion, and subsonic filtered to
suppress p-modes and residual jitter stemming from destretching
\citep{1989ApJ...336..475T}.

The final products of the reduction process were two time series 
($G$-band and $G$-continuum) for each one of the eight heliocentric 
positions on the solar disk, with a duration ranging from 28 to 100 images 
(from 5 to 17 min) as summarized in Table~\ref{ff_results}. 
They represent a set having homogeneous angular resolution and
covering a wide range of heliocentric angles.
Based on their contrast, we select the 
$G$-band image of best quality around the middle of each
time  series. They are the {\em reference images} used 
in our CLV analysis. Examples of two such images are shown
in Fig.~\ref{image}. 

\section{Center-to-limb variation of the solar surface 
occupied by bright points}\label{sec_ctl}

We have measured the fraction of solar surface occupied by BPs in 
each one of the eight $G$-band reference images described in 
\S~\ref{observation},
which correspond to different heliocentric positions on the disk. 
From now on this quantity will be referred to as 
{\em Fraction of Covered Surface} (FCS).

The detection of BPs has been performed by eye with the
help of computer tools. The procedure is the same used by 
\citet{2004ApJ...609L..91S,2010ApJ...715L..26S}, and 
can be briefly described as follows:  
the reference image of a particular time series is segmented with the algorithm
by \citet{1994PhDT.......347S} that, based of the sign of the second
derivative of the intensity at every image point, detects areas locally bright
(i.e. areas brighter than their surroundings).
The algorithm
creates a binary mask with the pixels of the bright patches 
set to 1.
Using an interactive program, we flick on the
computer display the reference image and the binary mask which allows us to
identify coincidences between $G$-band BPs and
small segmented patches. The BPs are selected one by one, considering as 
such those matching a segmented patch in the binary mask
when they overlay intergranular lanes and preserve identity
along several frames close to the reference image
(at least plus minus two frames).
We also select 
some faint bright features
that are not evident BPs in the reference image but 
which reveal themselves as such in the preceding or
the following image. As a general policy
doubtful structures are discarded. The procedure is  
carried out for two different saturation levels of the
reference image on the computer display --
low contrast to
select the brightest and more evident BPs, and high contrast to 
identify weaker structures.

The area occupied by the BPs is calculated as the number of
pixels corresponding to the selected BPs. The ratio between this area
and the total number of pixels in the image provides the FCS.
Thus, we get the FCS for every
$\mu$, as listed in Table~\ref{ff_results}. These values should be regarded as lower 
limits because of three reasons: (1) we only include 
secure BPs, (2) the number of detections increases with the 
angular resolution, which is finite, and (3) the segmentation algorithm
underestimates the area covered by the large BPs 
\citep{2010ApJ...715L..26S}. As an argument for consistency,
we point out that
the FCSs at disk center are in agreement 
with that obtained from images of similar quality using the same 
method by \citet[][$\sim\,$0.9\%]{2010ApJ...715L..26S}, 
but are higher than those obtained from lower quality 
images \citep[][$\sim\,$0.5\%]{2004ApJ...609L..91S}.

\begin{figure}[!h]
\centering
 \includegraphics[width=0.5\textwidth]{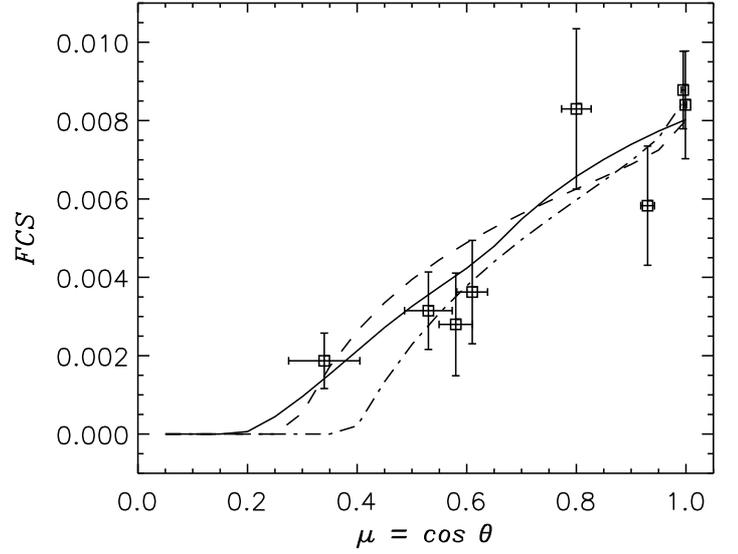}
\caption{Measurements of the fraction of solar surface
occupied by BPs, FCS, at 8 heliocentric positions.
Vertical bars represent the measure errors while horizontal 
bars show the
range of heliocentric positions included in the finite extension of the FOV. 
The three different curves represent the three different
solutions provided by our toy-model -- 
the solid, the dashed and the dot-dashed lines correspond to 
models a, b, and c in Table~\ref{fit1}, 
respectively.
}
\label{curvas8ptos}
\end{figure}
The CLV of the FCS values are displayed in Fig.~\ref{curvas8ptos}. 
Each symbol corresponds to one of the series, 
and includes error bars both in heliocentric angle and FCS.
The horizontal error bars represent the range of $\mu$ in 
the FOVs. The vertical error bars have been estimated 
dividing each full FOV into 24 non-overlapping subfields.
They represent the standard deviation of the 
mean FCS, considering the various FCSs in the 
independent sub-fields. These error bars estimate the statistical 
error of the measurement. They do not account for the systematic 
errors stemming from the subjectivity of our BP identification. 

%
%
\section{Interpretation}\label{interpretation}

\subsection{A toy-model kG magnetic concentration}\label{toy-model}


\begin{figure}[!h]
 \includegraphics[width=0.5\textwidth]{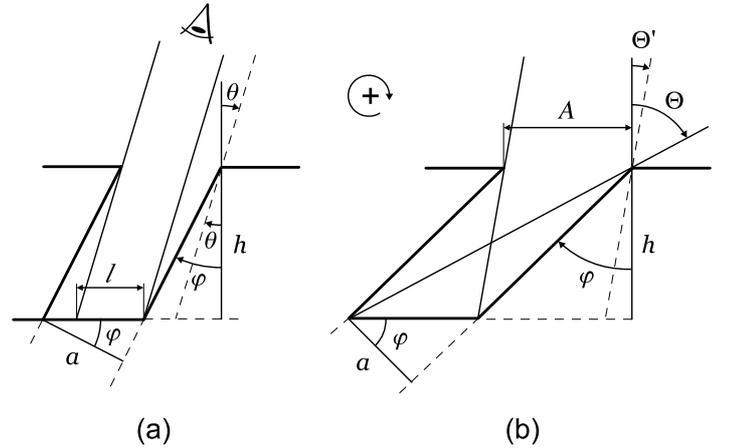}
\caption{(a) Vertical section of the toy-model magnetic
concentration used in the paper (the thick solid line). 
It represents a concentration of width $a$, 
tilted an angle $\varphi$ with respect to the local vertical 
and observed at heliocentric angle  $\theta$.
The symbols $h$ and $l$ are, respectively, the geometrical depth of the
concentration and the portion of the bottom accessible to observation.
(b) For a given tilt angle $\varphi$, $\Theta$ and $\Theta^\prime$
represent the extreme inclination angles
that allow us to observe 
the bottom of the magnetic concentration.
The encircled plus symbol between figures 
shows the positive
sense of rotation used in our equations.
}
\label{modelo}
\end{figure}

In essence, a kG magnetic concentration represents a depression
in the solar surface. Light comes from deeper photospheric 
layers which are usually hotter, and so the 
structure looks brighter. This basic 
idea, simplified to the extreme but retaining key ingredients, 
is considered in the 2-dimensional toy-model of magnetic 
concentration represented in Figure~\ref{modelo}a.
It portrays the vertical section of a magnetic concentration
of width $a$, which is located at an heliocentric angle $\theta$.
Due to the evacuation, we see deeper through the magnetic concentration,
and the difference of geometrical depths with respect to the 
non-magnetic photosphere is denoted by $h$.
The concentration can be inclined with respect to the
local vertical, an inclination that we parameterize as $\varphi$.
We assume the magnetic concentration to look bright only if 
our line-of-sight (LOS) reaches its bottom. 
In other words, the 
fraction of solar atmosphere that looks bright according to this 
model is proportional to $l$ (see Fig.~\ref{modelo}a).
For a given configuration of the concentration (set by 
$a,h,$ and $\varphi$), it shows up bright ($l\not= 0$) 
only for a limited range of heliocentric angles, i.e.,
$\Theta^\prime < \theta < \Theta$ as illustrated in 
Fig.~\ref{modelo}b. These extreme heliocentric angles 
are given by,
\begin{equation}
\tan \Theta = \tan \varphi + {a \over h \cos \varphi} ~~~ ; ~~~ \tan
\Theta^\prime = \tan \varphi - {a \over h \cos \varphi},
\label{eq:tube1}
\end{equation}
with the natural constrains $0 < \Theta < 90^\circ$ and $0 \leq \Theta^\prime <
\Theta$.
Given an heliocentric angle $\theta$ within the proper interval 
($\Theta^\prime$ ,$\Theta$),
the length of the  
concentration that is visible, $l$, turns out to be,
\begin{equation}
l = {a \over \cos \varphi} - h \mid \tan \theta - \tan \varphi \mid.
\label{eq:tube2}
\end{equation}
The FOV contains a number of these idealized magnetic
concentrations having different properties, therefore,
within this toy model, the FCS is proportional to 
the mean $l$ averaged over the set of magnetic 
concentrations, $\langle l\rangle$,
\begin{equation}
{\rm FCS=}f\,\langle l\rangle.
\label{model_fcs}
\end{equation}
The scaling factor $f$ accounts for the number of 
concentrations per unit surface, the area of the 
individual concentrations, and possibly other factors 
of order unity associated with the fact that the 
magnetic concentrations do have a 3-dimensional
structure not included in our 2-dimensional 
toy-model. In our case we additionally assume all 
concentrations to have the same $a$ and $h$, but 
different inclinations\footnote{We consider a 
distribution of tilts because the simpler 
alternative of assuming purely vertical magnetic 
concentrations 
produces a CLV with a sharp drop at small $\mu$, which is 
not observed.}, 
with the tilt $\varphi$ following 
a uniform distribution from $-\varphi_0$ to $+\varphi_0$ 
($\varphi_0 < 90^\circ$). 
Consequently, our model for the CLV of FCS depends of three 
parameters, namely, $h/a, \varphi_0$ and 
$af$. It does
not depend on $h$ and $a$ separately since the scaling 
factor $af$ absorbs the dependence on $a$ of $l$ 
(see equation~[\ref{model_fcs}]).
The closer to the limb, the higher in the
photosphere we observe \citep[e.g.][]{sti91}, a fact
that does not contradict our assumption of
$h/a$ to be independent of heliocentric angle. 
We are assuming that this systematic rising towards 
the limb affects the photosphere globally. In other 
words, both magnetic concentration and surroundings are
uplifted by the same amount, leaving unchanged 
the relative depth of the concentration $h$.

Obviously, it is difficult to justify the use of our 
toy-model based on its realism. The Sun is not 
2-dimensional, the magnetic concentrations are not slabs, 
their walls are also bright, the non-magnetic background 
is not uniform,
several concentrations may contribute 
along a single LOS, the depth $h$ depends on the field 
strength, and so on and so forth. However, the
model considers the basic physical ingredients responsible 
for the CLV  of the FCS, i.e., the bright structures are 
depressed with respect to the mean photosphere, so that 
they can be observed or not depending on their width, depth, 
and the relative orientation between their axes and the LOS.
Moreover, the use of the proper 3-dimentional
numerical models of magneto-convection is 
complicated,
and clearly goes beyond the scope of the work 
(see \S~\ref{intro}). Using models that consider 
3-dimensional magnetic flux-tubes fanning-out with height may be 
viewed as an intermediate alternative, but we disregarded this 
possibility from scratch since the analytic model become 
extremely involved,  and  its {\em realism} remains way off 
that of the proper numerical simulations of magneto-convection
(cf. 
\citeauthor{1993A&A...268..736B}~\citeyear{1993A&A...268..736B}
with 
\citeauthor{car04}~\citeyear{car04} or
\citeauthor{kel04}~\citeyear{kel04}).
The use of simplified models is also justified by the large 
scatter in the observed CLV 
(Fig.~\ref{curvas8ptos}), 
provided that they only intend to explain general trends. 

\subsection{Results of the model}\label{results}

As we pointed out in the previous section, the CLV in our
toy-model depends on only three independent
parameters, that we choose to be $h/a, \varphi_0$ and $af$. 
In order to determine which set of parameters reproduces 
the observation best, we carry out a least squares fit to determine the 
three free parameters. Since the model is non-linear,
the fit has to be carried out using an iterative procedure
for which we employ the standard Levenberg-Marquardt 
algorithm \citep[e.g.,][]{pre88}.  
Non-linear least squares fits do not grant 
uniqueness. The algorithm seeks and finds local minima 
of the merit function ($\chi^2$), and if it has
several, the best fitting parameters depend on 
the starting point. 
In order to get rid of this undesired 
dependence, we repeated the fit using $10^4$
different initializations, where the starting $h/a, \varphi_0$ and $af$
were obtained from three independent uniform distributions
spanning the full range of sensible values 
($0.2 \leq h/a \leq 4$, $0^\circ\leq \varphi_0 \leq 89^\circ$
and $5\times10^{-4} \leq af \le 5\times10^{-1}$).
The ensemble average needed to account the distribution
of tilts from $-\varphi_0$ to $\varphi_0$ (see \S~\ref{toy-model})
was evaluated numerically from $10^4$ samples drawn from a uniform 
distribution.
\begin{table*}                   
\caption{Parameters of the best fitting toy-model considering 
the eight heliocentric angles}   
\label{fit1}      
\centering                   
\begin{tabular}{c c c c c c r}          
\hline\hline                        
\noalign{\smallskip}
$\chi^2$ & $h/a$ & $\varphi_0$ [$^o$] & $af$ & $<A>$ & solutions & [$\%$] \\   
\hline
\noalign{\smallskip}
0.90--0.91 & $0.65\pm0.16$ & $64\pm 5$ & $(9.7\pm1.6)\times10^{-3}$ & 65.9 & a & 72.8 \\
1.45--2.00 & $0.29\pm0.01$ & $ 1\pm44$ & $(8.0\pm0.3)\times10^{-3}$ & 50.0 & b & 23.5 \\
1.97--1.98 & $0.43\pm0.02$ & $ 6\pm24$ & $(8.8\pm0.3)\times10^{-3}$ & 50.1 & c & 2.8 \\
\noalign{\smallskip}
\hline                                   
\end{tabular}
\end{table*}
Essentially, the different initializations converge to three different solutions. 
Table~\ref{fit1} summarizes the parameters characteristic
of these three families -- it gives the mean values among all 
solutions in the family plus the formal error bars 
provided by the $\chi^2$-minimization algorithm. 
The corresponding fits are plotted in Fig.~\ref{curvas8ptos}.
The most recurrent result (73\,\%\ of the accepted solutions -- 
a in Table~\ref{fit1}) is also the most reliable since it represents 
the smallest $\chi^2$ close to $1$ (Table~\ref{fit1}). It corresponds
to fairly shallow magnetic concentrations (aspect ratio $h/a\approx 3/5$)
with a large range of tilts ($\varphi_0 \approx 64^\circ$). The 
second most common solution (b in Table~\ref{fit1}) 
also corresponds to shallow 
magnetic concentrations ($h/a\approx 3/10$) but they are vertical.
Its $\chi^2$ is slightly larger than that for solution a. 
The third case is intermediate between a and b.

\begin{figure}
\includegraphics[width=0.5\textwidth]{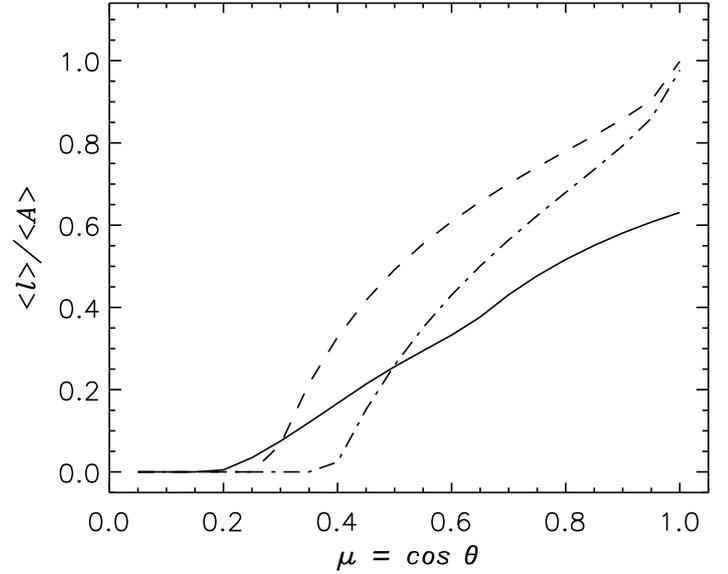}
\caption{Fraction of magnetized surface 
producing BPs. The three curves correspond
to the predictions of our toy-model for the three 
solutions portrayed in Fig.~\ref{curvas8ptos}. Note that 
magnetic structures no longer render BPs as one 
approaches the limb.}
\label{true_detection}
\end{figure}

According to the model and its assumptions, one just 
{\em sees} 
as BPs a portion $l$ of the surface, which is only 
a fraction of the true surface occupied 
by magnetic structures $A$ -- see Fig.~\ref{modelo}. The 
two quantities agree when the LOS coincides with the axis of 
the concentration, but in general $A\geq l$. 
In our toy-model the fraction of true surface that is 
observed as BP turns out to be   $\langle l\rangle /\langle A\rangle$, 
where the brackets account for the fact that we consider
an ensemble of magnetic concentrations. The variation with 
heliocentric angle of this fraction is represented 
in Fig.~\ref{true_detection}. 
As one approaches the limb, most
of the magnetic concentrations are no longer BPs -- 
less than 20\,\% of them at $\mu < 0.5$.
Figure~\ref{true_detection} also shows how the solution with large range
of magnetic field inclinations (a 
in Table~\ref{fit1}, and the solid line in 
Fig.~\ref{curvas8ptos}) 
hides a large fraction of the existing magnetic structures; 
only 60\% of them are observed as BPs at 
disk center. 

\begin{figure}
 \includegraphics[width=0.5\textwidth]{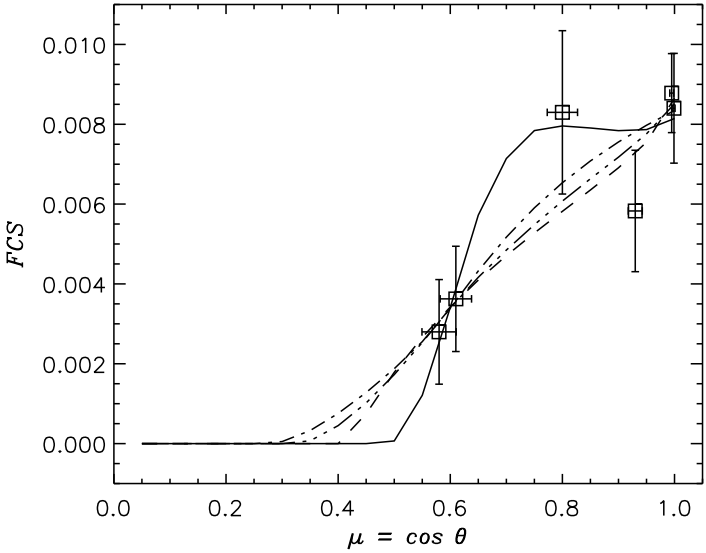}
\caption{Same as Fig.~\ref{curvas8ptos} but for fits to the six 
center-most data points
describing the CLV. The solid line, the dashed line, the dot dashed line, 
and the triple-dot dashed line correspond to solutions d, e, f and
g in Table~\ref{fit2}, respectively.
}
\label{curvas6ptos}
\end{figure}

The closer to the limb the more difficult the BP identification is
(\S~\ref{observation} and Fig.~\ref{image}). In order to explore
the effect of this uncertainty, we repeated the fits removing the two FOVs 
closest to the limb ($\mu=\,$ 0.34 and 0.53).
The results are shown in Table~\ref{fit2} and Fig.~\ref{curvas6ptos}.
Table~\ref{fit2} reflects the presence of four families of solutions, 
all of them with similar $\chi^2$. The third most
frequent result (f in Table~\ref{fit2} -- 17 $\%$ of cases) depicts a
shallow flux tube $h/a \approx 3/5$ with a significant tilting range 
$\varphi_0 \approx 46^\circ$, i.e., it looks quite similar to the main 
solution a in Table~\ref{fit1}. However, the
most common solution (40 $\%$ of cases -- d in Table~\ref{fit2}) 
represents a much deeper tube, $h/a > 3$, although with a similar tilting 
range ($\varphi_0 \approx 52^\circ$). Finally, 
the solution e in Table~\ref{fit2}
resembles very much to that reported as solution c in Table~\ref{fit1}.
Removing the two most uncertain FCSs yields solutions similar to
the full case, but it also introduces another possibility of 
rather slim magnetic concentrations (d in Table~\ref{fit2}).

\begin{table*}                   
\caption{Parameters of the best fitting toy-model considering 
the six center-most fields}   
\label{fit2}      
\centering                   
\begin{tabular}{c c c c c c c}          
\hline\hline                        
\noalign{\smallskip}
$\chi^2$ & $h/a$ & $\varphi_0$ [$^o$] & $af$ & $<A>$ & solutions & [$\%$] \\   
\hline
\noalign{\smallskip}
0.47 &      $3.3\pm1.6$ & $52\pm 1$ & $(4.7\pm2.4)\times10^{-2}$ & 59.0 & d & 39.5 \\
0.51--0.52 & $0.46\pm0.04$ & $ 7\pm29$ & $(8.9\pm0.4)\times10^{-3}$ & 50.1 & e & 29.9 \\
0.54 &      $0.59\pm0.24$ & $ 46\pm33$ & $(9.7\pm1.6)\times10^{-3}$ & 56.6 & f & 17.2 \\
0.52 &      $1.11\pm0.36$ & $ 54\pm5$ & $(1.5\pm0.5)\times10^{-2}$ & 59.8 & g & 10.9 \\
\noalign{\smallskip}
\hline                                   
\end{tabular}
\end{table*}

In short, the large scatter of the observations makes it 
easy to reproduce the CLV  
with our toy-model. The most
favored solution corresponds to BPs being rather shallow
structures with a large spread of vertical inclinations.
This solution is not unique,  and the model also allow for
predominantly vertical structures and for narrow 
concentrations.

\section{Fraction of solar disk covered by BPs}\label{tsi_var}

\begin{figure}
\includegraphics[width=0.4\textwidth]{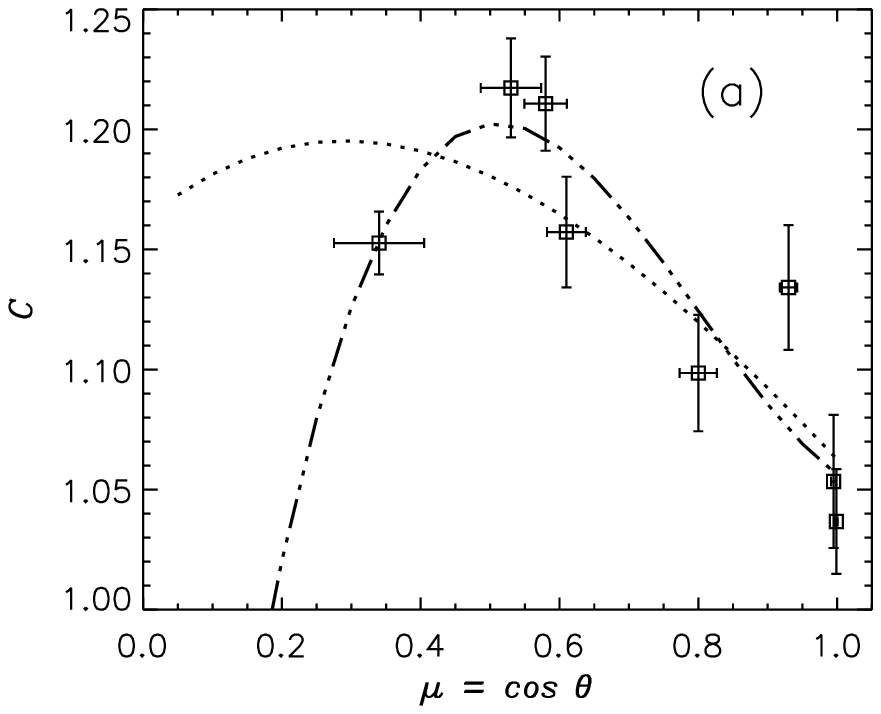}
\includegraphics[width=0.4\textwidth]{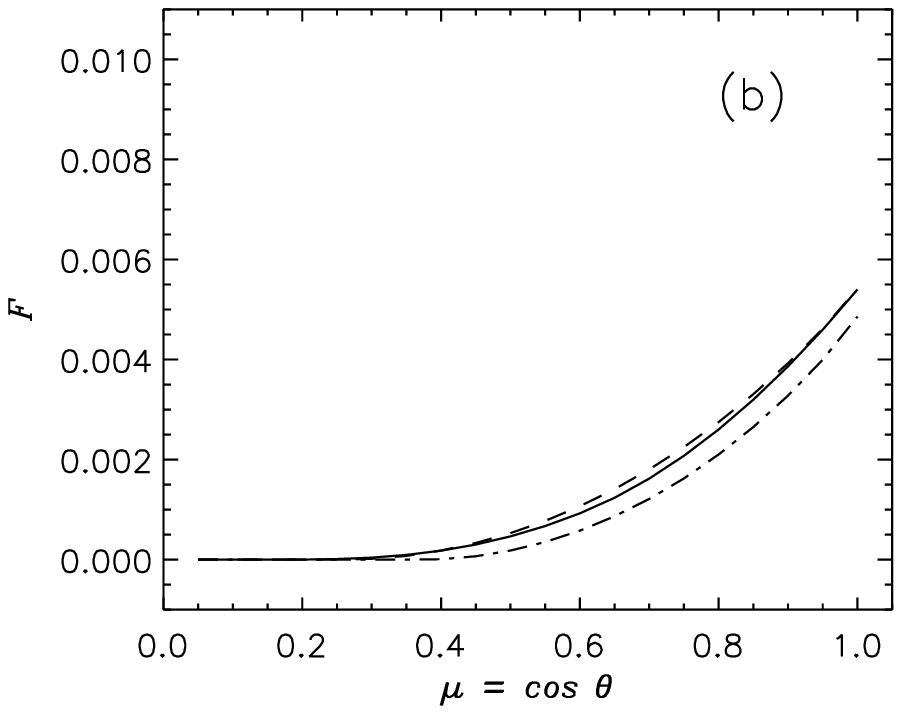}
\includegraphics[width=0.4\textwidth]{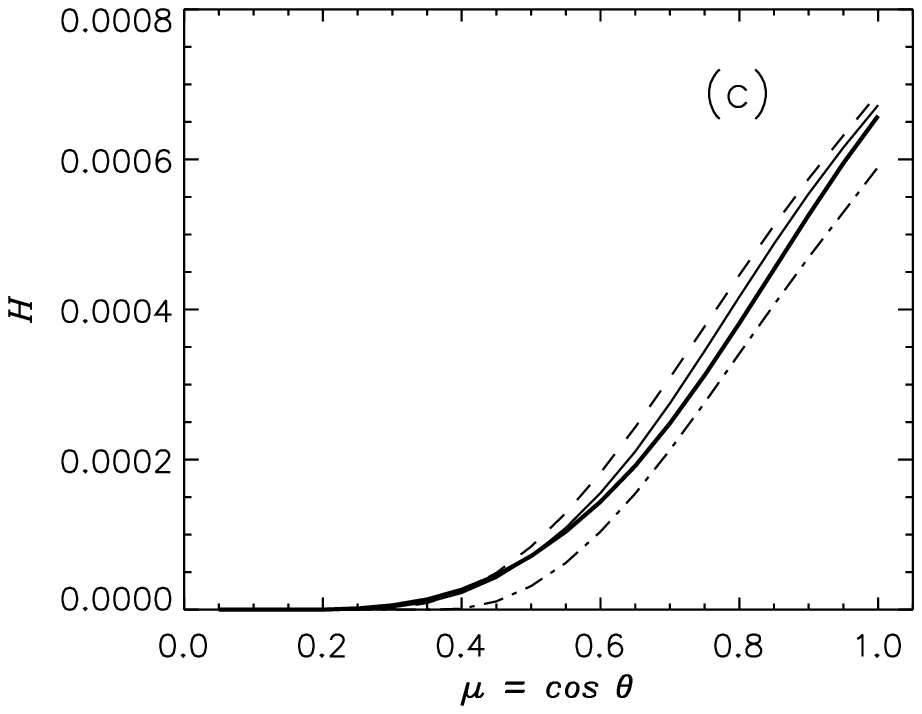}
\caption{
(a) 
Contrast of the BPs in the $G$-band as observed in our fields,
$C$. The
symbols are observations, and the lines represent two
different smooth fits to the observed CLV.
(b) Fraction of total solar disk covered by BPs 
from the solar limb to a particular $\mu$ 
-- $F$ in equation~(\ref{auxiliary}).
The three different types of line correspond to the
three curves in Fig.~\ref{curvas8ptos}.
(c) Fraction of TSI contributed by the 
quiet Sun BPs
-- $H$ in equation~(\ref{irrad}).
The three different types of thin lines correspond to the
three curves in Fig.~\ref{curvas8ptos}, with 
the contrast given by the triple-dot dashed line in (a).
The thick solid line is equivalent to the thin solid line
except that the contrast is taken as the dotted line in (a). 
} 
\label{integrated_area}
\end{figure}

${\rm FCS}(\mu)$ describes the fraction 
of surface covered by BPs at each location
on the Sun. Once ${\rm FCS}(\mu)$ is known, 
the fraction of the solar disk covered by BPs 
can be determined by integration. The fraction of solar disk 
with heliocentric angle between $\mu$ and $\mu+d\mu$ 
is $2\mu\,d\mu$ \citep[e.g.,][]{1991ApJ...383L..89F}, 
therefore,
\begin{equation}
F(\mu)=2\int_0^\mu {\rm FCS}(\mu')\,\mu'\,d\mu',
\label{auxiliary}
\end{equation}
provides the fraction of surface with BPs 
from the limb to $\mu$. Obviously, $F(1)$ represents the 
fraction covering the full Sun.
$F(\mu)$ is given in Fig.~\ref{integrated_area}b
for the three solutions in Fig.~\ref{curvas8ptos}.
The three functions are very similar with
$F(1)\simeq 0.005$, i.e., some 0.5\,\% of the solar disk is 
covered by BPs. The disk center is the maximum contributor 
to this BP covering -- 90\,\% of it is due BPs at $\mu > 0.5$, 
and half of the covering is associated with BPs at  
$\mu > 0.8$ (Fig.~\ref{integrated_area}b). Even though the above 
conclusions use the CLV of our toy-model, it is important
to emphasize that they are independent of  
the assumptions we made to work it out. 
The model CLVs are used here as smooth continuous 
approximations to describe the actual observations.

As we argue in the previous paragraph, most of the area 
occupied by BPs is at the disk center. It seems that any  
potential contribution of the quiet Sun magnetic fields 
to the TSI cannot come from the solar limb. In order to make 
this statement more quantitative, we follow the 
equations by \citet{1991ApJ...383L..89F} to write down 
the contribution to the TSI of the BPs, i.e., it is
given as 
\begin{displaymath}
\frac{\Delta S}{S}=H(1),
\end{displaymath}  
with
\begin{equation}
H(\mu)=\frac{1}{2}\int_0^\mu{\rm FCS}(\mu')\,\big[C(\mu')-1\big]\,
(3\mu'+2)\,\mu'\,d\mu'.
\label{irrad}
\end{equation}
The symbol $\Delta S$ stands for the change on the solar irradiance 
$S$ produced by the quiet Sun BPs. The previous equation
assumes an Eddington limb-darkening for the quiet Sun intensity. 
The parameter $C(\mu)$ is the contrast of the BPs relative to
the mean intensity averaged over all wavelengths. 
(Just to make the meaning of $\Delta S/S$ more clear:
if BPs twice as bright as the mean photosphere fully
cover the disk, i.e. $C=2~\forall~\mu$, then  
$\Delta S/S=1$.)  As we explain above, 
the part of the integrand of equation~(\ref{irrad}) 
involving {\rm FCS} is strongly biased towards
the disk center. The only parameter that may counter-balance
this effect is $C$. If $C$ increases too rapidly
with $\mu$ it may dominate the integrand, and so, the TSI.
We ignore how $C$ varies from center to limb, however,
as an educated-guess to evaluate $H(\mu)$,
we have assumed $C$ to scale as the intensity 
variation we measure in the $G$-band.  This CLV of the contrast 
in the $G$-band is given in Table~\ref{ff_results} 
and corresponds to the symbols represented in 
Fig.~\ref{integrated_area}a. Two smooth 
curves fitted to the observed points are also included 
in the figure -- they consider the two extreme 
scenarios of the contrast remaining constant at the limb 
(the dotted line) or dropping  down significantly   
(the triple-dot dashed line). 

Figure~\ref{integrated_area}c displays 
$H(\mu)$ for the three solutions in Fig.~\ref{curvas8ptos} 
considering the contrast decreasing at the limb.
They are very similar showing in all cases that 
the irradiance is strongly biased towards the disk 
center. This result is independent on whether the 
CLV of $C$ has or not the drop at the limb -- the thick
solid line in Fig.~\ref{integrated_area}c is equivalent
to the thin solid line except that the used $C$ maintains
a constant value at the limb (the dotted line in  
Fig.~\ref{integrated_area}a). From $H(\mu)$ in 
Fig.~\ref{integrated_area}c one finds that
90\,\% of the TSI is contributed by BPs having $\mu > 0.5$, 
and some half of it is due to BPs with $\mu > 0.8$.

The peak value of $H$ is about 0.07\,\% (Fig.~\ref{integrated_area}c).
It corresponds to BPs with a typical contrast of the order of 1.15 
filling the observed disk coverage of some
0.5\,\% (Fig.~\ref{integrated_area}b).
This figure for the 
contribution of the quiet Sun magnetic fields to TSI is similar 
to the recent estimate by \citet{2011A&A...532A.136S} mentioned 
in \S~\ref{intro}. Actually, it is a factor of two smaller, 
but their estimate refers to the disk center observed at a 
different wavelength. In view of the uncertainties involved 
in this type of work, the coincidence is worth pointing out.
The two estimates are completely different and yet
provide consistent results. 
   
\section{Discussion and Conclusions}\label{conclusions}

We have measured the center-to-limb variation (CLV)
of the area covered by $G$-band bright points (BPs)
in the quiet Sun (fraction of covered surface or 
FCS). It is a parameter difficult to determine since the 
detection of BPs critically depends on the angular 
resolution of the observation (\S~\ref{intro}). We employ 
several time series taken in 
two hours
during moments of excellent seeing with the SST.
The images were post processed using MOMFBD (see
\S~\ref{observation}) which provides a homogeneous
set of images adequate for these subtle measurements.
They were restored to provide an angular resolution close
to the diffraction limit of the instrument at the working
wavelength (some 0\farcs 1 at the $G$-band).

We find the FCS to be largest at disk center ($\simeq$ 1\,\%),
and then it drops down to become $\simeq$ 0.2\,\% at $\mu\simeq 0.3$.
The relationship has large scatter, which we managed to 
estimate comparing different subfields within our FOVs
(see the error bars in Fig.~\ref{curvas8ptos} and 
Table~\ref{ff_results}). The value obtained at the disk center 
agrees with previous estimates based on data of similar 
quality. 

We work out a toy model to describe the observed CLV.
It assumes the magnetic bright points to be depressions
in the mean solar photosphere, characterized by a depth,  
a width, and with a spread of inclinations. It is only an 
exploratory modeling which, however, includes the 
physical ingredients that seems to be responsible
for a kG magnetic concentration to show up 
as a bright feature on the solar disk. The solutions
offered by our toy-model are poorly constrained, but they
seem to show the BPs to be shallow 
structures\footnote{This result is not inconsistent with
the magnetic concentrations producing BPs being modeled
as a compactly-packed ensemble of narrow magnetic 
concentrations arranged in a micro-structured magnetic 
atmosphere MISMA fashion \citep[][]{1996ApJ...466..537S,
2000ApJ...544.1135S,2001ApJ...555..978S}.
The CLV of the FCS is sensitive to the global scale of the ensemble, 
whereas the MISMA accounts for the smallest scales 
responsible, among others, for the asymmetries of the 
spectral lines formed in the atmosphere.
}
(ratio depth to width $\simeq 0.7\pm 0.2$) with 
a large spread of inclinations  ($\simeq \pm 70^\circ$).

Among others, the FCS is of interest because it  
determines the impact that quiet Sun magnetic fields may have on 
TSI variations, an influence so far unknown. Since the measured 
FCS is so peaked towards disk center, any role that quiet Sun 
magnetic fields may have to play on TSI will be due to the center 
of the disk. According to our estimate, 90\,\% of the TSI is contributed 
by BPs having $\mu > 0.5$,  and half of it is due to BPs 
with $\mu > 0.8$ (\S~\ref{tsi_var}). This estimate is based on 
assuming the CLV of the BP (wavelength integrated) flux
to scale as the observed $G$-band intensity. It is an ad-hoc 
assumption  adopted for lacking of a better one which, however,
has allowed us to provide a first constrain on the effect of 
quiet Sun BPs on TSI.

In this sense, we have to stress that the FCS 
has been measured in the $G$-band, because the magnetic 
concentrations are particularly conspicuous at this 
wavelengths.Then we implicitly assume throughout the work 
that the FCS is the same at all wavelengths, i.e., that the 
area covered by magnetic concentration does not change with 
wavelength. (The $G$-band BPs are brighter but not bigger.)
This assumption remains to be proven however, 
for the time being, it seems to be a reasonable working 
hypothesis.

Our toy-model suggests the BPs to be shallow structures with varied
inclinations. One could test these predictions observing polarization signals of
quiet Sun BPs at disk center. On the one hand,
the polarization signals provide magnetic field
inclinations through Stokes inversion \citep[e.g.,][]{2008ApJ...674..596S}.
On the other hand, assuming the magnetic concentrations 
to be in mechanical balance, one can infer their depths 
\citep[e.g.,][]{2000ApJ...532.1215S}. 
These inferences require low-noise spectro-polarimetry with an angular 
resolution similar to that of our $G-$band images. 
They represent a technical challenge, but such 
observations seem to be doable in the near future 
\citep[see, e.g.,][]{2010ApJ...723L.164L}. 
As we mentioned in the introduction, the  existing numerical 
simulations of magneto-convection predict the BPs to be depressed
with respect to the mean photosphere \citep[e.g.,][]{vog05,vog07}.
Whether the faint BPs observed in quiet Sun are predicted to
be superficial or deep remains to be worked out. However, 
the existing simulations of plage magnetic concentrations 
suggest the continuum intensity to be formed in a rather
shallow region \citep{kel04,car04}.

-------------------------------


\begin{acknowledgements}

This work has been partially funded by the Spanish Ministry of Science and
Innovation through projects AYA2010--18029, ESP2006-13030-C06, ESP2003-07735-C04-04 
and AYA2009-14105-C06. Financial support by the European Commission through the 
SOLAIRE Network (MTRN-CT-2006-035484) is gratefully acknowledged.
The SST is operated in the Spanish Observatorio del Roque de los Muchachos  
by the Institute for Solar Physics of the Royal Swedish Academy of 
Sciences.

\end{acknowledgements}


\end{document}